\documentclass[pdflatex,sn-mathphys-num]{sn-jnl}

\usepackage[T1]{fontenc}

\usepackage{graphicx}%
\usepackage{multirow}%
\usepackage{amsmath,amssymb,amsfonts}%
\usepackage{amsthm}%
\usepackage{mathrsfs}%
\usepackage[title]{appendix}%
\usepackage{xcolor}%
\usepackage{textcomp}%
\usepackage{manyfoot}%
\usepackage{booktabs}%
\usepackage{algorithm}%
\usepackage{algorithmicx}%
\usepackage{algpseudocode}%
\usepackage{listings}%
\usepackage{xspace}
\usepackage{hyperref}
\usepackage{cleveref}
\newcommand{\MZ}{$\text{M}_\text{Z}$\xspace}
\newcommand{\MW}{$\text{M}_\text{W}$\xspace}
\newcommand{\MT}{$\text{M}_\text{t}$\xspace}
\newcommand{\NNNLO}{$\text{N}^3\text{LO}$\xspace}
\newcommand{\MCatNLO}{M\protect\scalebox{0.8}{C}@N\protect\scalebox{0.8}{LO}\xspace}

\newcommand{\POWHEG}{P\protect\scalebox{0.8}{OWHEG}\xspace}

\newcommand{\NLOPS}{N\scalebox{0.8}{LO}P\scalebox{0.8}{S}\xspace}
\newcommand{\NNLOPS}{NN\scalebox{0.8}{LO}P\scalebox{0.8}{S}\xspace}

\newcommand{\UNNLOPS}{UNN\scalebox{0.8}{LO}P\scalebox{0.8}{S}\xspace}

\newcommand{\MINNLOPS}{MINN\protect\scalebox{0.8}{LOPS}\xspace}
\newcommand{\Geneva}{G\protect\scalebox{0.9}{eneva}\xspace}
\newcommand{\Sherpa}{S\protect\scalebox{0.9}{HERPA}\xspace}

\theoremstyle{thmstyleone}%
\theoremstyle{thmstyletwo}%

\theoremstyle{thmstylethree}%

\begin{document}

\title[Article Title]{Monte Carlo Event Generators for Future Lepton Colliders
\footnote{Invited contribution to the International Workshop on Future Linear Colliders (LCWS 2025), 20-24 October 2025. Valencia, Spain (C25-10-20.1)}}


\author[1]{\fnm{Alan} \sur{Price}}\email{alan.price@uj.edu.pl}



\affil[1]{Jagiellonian University, ul.\ prof.\ Stanis\l{}awa \L{}ojasiewicza 11, 30-348 Krak\'{o}w, Poland}



\abstract{
    Monte Carlo event generators are essential tools in collider physics, providing the link between 
    theoretical predictions and experimental measurements through fully exclusive event simulation. 
    Future collider programmes, particularly high-precision lepton colliders will place significantly increased demands on their accuracy and scope.
    This contribution reviews key challenges in MC generator development, including electroweak corrections, 
    initial-state radiation, beam dynamics, perturbative QCD, and non-perturbative modelling. 
    The discussion is not exhaustive and reflects a selective choice of topics.
}

\maketitle

\section{Introduction}\label{sec1}
Monte Carlo (MC) event
generators~\cite{Sherpa:2019gpd,Sherpa:2024mfk,Sjostrand:2006za,Sjostrand:2014zea,Bierlich:2022pfr,Alwall:2011uj,Alwall:2014hca,Bahr:2008pv,Bellm:2015jjp,Bewick:2023tfi,Kilian:2007gr,Jadach:2022mbe,Jadach:1999vf,CarloniCalame:2003yt,Alioli:2010xd}
occupy a unique and indispensable position in high-energy physics. They serve
simultaneously as precision numerical frameworks, phenomenological models of
non-perturbative dynamics, and the essential interface between theoretical
predictions and experimental measurements. In practice, they are the only tools
capable of providing fully exclusive final states that can be directly compared
to detector-level observations. As such, they underpin essentially every stage
of collider physics, from detector design and performance studies to precision
measurements and searches for new phenomena. 

At the Large Hadron Collider (LHC), decades of sustained development have led
to a remarkably mature and sophisticated ecosystem of MC
generators~\cite{Buckley:2011ms,Campbell:2022qmc,Reuter:2025imz}. Advances in
perturbative calculations, such as automated algorithms for NLO
calculations~\cite{Catani:1996vz,Frixione:1995ms},
tree-level~\cite{Stelzer:1994ta,Caravaglios:1995cd,Krauss:2001iv,Gleisberg:2008fv,Moretti:2001zz,Kanaki:2000ey,Mangano:2002ea}
and
one-loop~\cite{Denner:2005nn,Ossola:2007ax,Peraro:2014cba,Denner:2014gla,Ossola:2006us,GoSam:2014iqq,Hirschi:2011pa,Buccioni:2019sur,Actis:2016mpe,Berger:2008sj}
amplitudes, and their consistent matching and merging with parton
showers~\cite{Frixione:2007vw,Frixione:2002ik,Monni:2019whf,Jadach:2015mza,Hoeche:2005vzu,Catani:2001cc,Hamilton:2012np,Herren:2022jej,Hoche:2025gsb,vanBeekveld:2025lpz}
have been complemented by increasingly refined models of the non-perturbative
regime, such as underlying event~\cite{Cacciari:2007fd,Sjostrand:1987su} and
hadronization~\cite{Andersson:1983ia,Amati:1979fg}. These developments have
been tightly constrained and iteratively improved through detailed comparisons
to experimental data across a wide range of energies and
processes~\cite{Iskauskas:2026rxi,Buckley:2009bj,LaCagnina:2023yvi,Chahal:2022rid}.
The resulting level of reliability is often taken for granted today, but it
reflects a long and non-trivial co-evolution of theory, phenomenology, and
experiment, supported by continuous validation efforts and large-scale tuning
campaigns. In recent years there has also been considerable effort devoted to
extending the perturbative accuracy to next-to-next-to-leading order (NNLO) in
QCD~\cite{Gehrmann-DeRidder:2005btv,Catani:2007vq,Czakon:2010td,Boughezal:2011jf,Currie:2013vh,Cacciari:2015jma,Gaunt:2015pea,Magnea:2018hab,Caola:2017dug,Grazzini:2017mhc}.
This progress has been driven by the need to match the increasing precision of
experimental measurements, and has led to the development of several approaches
for consistently combining NNLO calculations with parton
showers~\cite{Alioli:2013hqa,Hoche:2014uhw,Campbell:2021svd,Karlberg:2014qua},
thereby enabling fully exclusive event generation at this level of accuracy.

In addition to providing predictions for the Standard Model (SM), MC tools are 
an essential tool for simulations of physics beyond the SM (BSM). Such simulations 
require a pipeline that spans the definition of a new model's Lagrangian, the derivation 
of Feynman rules, the generation of hard-process matrix elements, and their interface 
with the full generator chain including parton showers and hadronization. Historically, 
each new BSM model would need to be implemented on a case-by-case basis, requiring substantial 
manual effort at every stage. However, the advent of automated model-building and interface 
tools has drastically streamlined this workflow. Spectrum generators and model derivation 
tools such as LanHEP~\cite{Semenov:2008jy}, SARAH~\cite{Staub:2013tta}, and Feynrules~\cite{Alloul:2013bka}
 allow Lagrangians 
to be specified in a compact form. The resulting model information 
can then be exported in the Universal FeynRules Output (UFO) format~\cite{Degrande:2011ua,Darme:2023jdn} 
or the FeynRules model format, providing 
a standardised interface to generators~
\cite{Frixione:2019fxg, Reuter:2024dvz,Stienemeier:2021pux,Degrande:2020evl,Hoeche:2014rya,CONTUR:2021qmv}. 
Public model libraries such as the FeynRules model database 
 further facilitate the dissemination and validation of BSM implementations across the community. 
 At the level of the hard process, BSM simulations benefit from the same NLO automation 
 machinery developed for SM processes, including dedicated NLO BSM calculations matched 
 to parton showers~\cite{Degrande:2014vpa,Hirschi:2015iia}. The combination of these tools 
 means that a phenomenologically complete simulation of a new BSM scenario, from model 
 definition through to fully hadronised final states, can now often be achieved with modest effort.

Looking ahead, future collider programmes~\cite{Altmann:2025feg} will require a
comparable, if not significantly higher level of generator sophistication,
while simultaneously introducing qualitatively new challenges. At
high-precision lepton colliders such as CEPC~\cite{CEPCStudyGroup:2018ghi},
CLIC~\cite{Aicheler:2012bya},FCC-ee~\cite{FCC:2018evy}, and
ILC~\cite{ILC:2013jhg}, theoretical uncertainties must be reduced to a level
commensurate with experimental accuracies that in some cases surpass those
achieved at LEP by an order of magnitude~\cite{Blondel:2019qlh}. This places
stringent demands on higher-order calculations, resummation techniques, and the
control of electroweak and QED effects within event generators.

In parallel, the computational demands associated with MC event generation are
expected to grow
dramatically~\cite{HEPSoftwareFoundation:2025irs,HEPSoftwareFoundation:2017ggl}.
The scale of future experimental programmes will require unprecedented volumes
of simulated data, both for precision measurements and for the training of
modern analysis techniques. This raises pressing questions about algorithmic
efficiency, parallelisation, and the integration of new computational
paradigms, including hardware acceleration and machine learning, into the
generator
workflow~\cite{Heimel:2022wyj,Heimel:2023ngj,Bothmann:2020ywa,Brass:2018xbv,
Carrazza:2020rdn,Hagiwara:2026lul,Heimel:2026hgp,Frederix:2026ejl,Bothmann:2025lwg,Herrmann:2025nnz,Villadamigo:2025our}.


\section{Physics demands of future lepton colliders}
Lepton colliders offer a qualitatively different environment from hadron
machines: a well-defined initial state, no hadronic remnants, and the ability
to operate at precisely controlled centre-of-mass energies. These advantages
translate directly into precision physics opportunities, but they also impose
stringent demands on theoretical predictions and, consequently, on Monte Carlo
generators. The machines currently under consideration, CEPC, CLIC, FCC-ee, and
ILC, span a wide range in energy, luminosity, and beam technology, but share a
common requirement for generator accuracy that in several areas substantially
exceeds what was needed at LEP.

The most immediate and perhaps most demanding context will be the unprecedented
precision measurements of electroweak observables~\cite{Freitas:2019bre}. For
example, FCC-ee's projected dataset at the Z pole, $\mathcal{O}(10^{12})$ Z
decays compared to $\mathcal{O}(10^7)$ at LEP, transforms what were once
theory-limited observables into measurements of unprecedented sensitivity. The
anticipated precision on \MZ is of order 0.1 MeV, roughly twenty times better
than the current LEP average. At the WW threshold, FCC-ee aims to determine \MW
through a cross-section scan with a precision of a few MeV, while ILC at 250
GeV and 500 GeV targets Higgs coupling measurements at the sub-percent level
through $e^+e^- \to ZH $ and WW fusion. In all of these cases, the extraction
of the physics quantity of interest depends critically on the accuracy with
which initial and final state radiation, the line shape of resonances, and the
interference between production and decay processes are modelled in the
generator. These are not marginal corrections: at FCC-ee, the QED radiator
function alone shifts the effective hadronic cross section at the Z pole by
tens of percent, and its higher-order terms contribute at a level comparable to
the target experimental precision.

Multi-boson production and the associated treatment of unstable particles
present a related set of demands that run across all three machines, but with
varying emphasis. At FCC-ee's WW threshold, the doubly-resonant $e^+e^- \to
    W^+W^- \to 4f$~\cite{Denner:2000bj,Jadach:1997llk} process must be described
including finite-width effects, non-factorisable corrections connecting
radiation from the two W decay products, and the full gauge-invariant set of
diagrams contributing to the four-fermion final state, effects which at the
permille level of precision targeted by FCC-ee cannot be approximated by the
naive narrow-width product of on-shell cross sections and branching fractions.
At ILC energies, WW fusion and its interference with ZZ fusion contribute to
the same Higgs final states, requiring a careful treatment of the full set of
contributing diagrams rather than a simple sum of factorised contributions. At
CLIC's higher energy stages, processes that are formally subleading at lower
energies, six-fermion final states, triple gauge boson production, WW
scattering, appear at leading order in the relevant signal and background
channels, demanding multi-leg generator capabilities across a wide range of
multiplicities and final-state topologies.

The top-quark threshold~\cite{Gusken:1985nf,Vos:2016til,Hoang:2000yr} is a
physics target for FCC-ee and, in a different kinematic regime, to
CLIC~\cite{CLICdp:2018esa}, and a possible upgraded CEPC, it presents its own
generator challenges. FCC-ee aims for a top-quark mass measurement at the
$t\bar{t}$ threshold with a precision of order 10–20
MeV~\cite{Simon:2016htt,Defranchis:2025auz}, exploiting the strong sensitivity
of the threshold lineshape to the mass in a well-defined short-distance scheme.
The accurate simulation of this lineshape requires the combination of
non-relativistic QCD resummation in the threshold
region~\cite{Beneke:2015kwa,Hoang:2013uda,Martinez:2002st,Beneke:2024sfa} with
a complete description of top production and decay, including QCD corrections
to both vertices, finite-width effects, and the Coulomb resummation of the
$1/v$ singularity near threshold~\cite{Bach:2017ggt}.

Beam effects introduce an additional layer of physics that is specific to
lepton colliders but varies significantly between machines. At FCC-ee, the
dominant effect is a precisely characterised beam energy spread that must be
convolved with the sharp threshold dependences exploited in \MZ, \MW, and \MT
measurements; its implementation is well-understood but must be propagated
consistently through the generator chain. At ILC and CLIC, beamstrahlung,
synchrotron radiation emitted by one beam in the electromagnetic field of the
opposing bunch, produces a luminosity spectrum with a long low-energy tail and
a peak at the nominal centre-of-mass energy, whose shape depends on the machine
operating parameters. This spectrum must be accurately modelled and convolved
with the hard process cross section, and at CLIC energies the two-photon
background from beamstrahlung photon interactions must be overlaid on hard
events in a manner analogous to pile-up at a hadron collider. Beamstrahlung
grows more severe with beam energy and density, and at CLIC's 3 TeV stage it is
a significant systematic for any measurement that uses the beam energy to
constrain event kinematics. ILC's longitudinal beam polarisation, available for
both electrons and positrons, adds a further requirement: the full
spin-correlated description of production and decay, including the polarisation
dependence of higher-order corrections, must be maintained throughout the
generator chain.

At the highest energies foreseen for CLIC, qualitatively new perturbative
effects come into play. Electroweak Sudakov
logarithms~\cite{Denner:2024yut,Pagani:2021vyk,Bothmann:2021led,Denner:2000jv,Ciafaloni:2000df,Ciafaloni:1998xg,Lindert:2026ief,Fadin:1999bq,Kuhn:1999nn,Lindert:2023fcu}
of the form $\alpha \log^2(s/m_W^2)$ grow to tens of percent at 3 TeV and must
be resummed or computed to high fixed order to obtain reliable predictions for
precision observables and background estimates alike. The hierarchy between QCD
and EW corrections, which at lower energies allows a natural ordering of the
perturbative expansion, becomes less clear-cut when both are large; consistent
schemes for their simultaneous treatment in a shower framework are still under
development. And if heavy new physics states exist and are produced at CLIC,
the cascade decays and complex multi-particle final states that would follow
require reliable background modelling in a regime where multiple hard scales
are simultaneously present.

A thread running through all of these contexts is that the generator demands
arise directly and unavoidably from the core physics goals of each machine.
They cannot be sidestepped by analysis choices or absorbed into experimental
systematic uncertainties. Achieving the precision targets of future lepton
colliders requires that the generators themselves be built to a correspondingly
higher standard, which means the development work must begin substantially
before the machines operate.

\section{Beam Dynamics}
At high-energy lepton colliders, the properties of the colliding beams
themselves play a critical role in shaping the effective collision energy and
event kinematics. Two dominant phenomena, beamstrahlung and the intrinsic beam
energy spread~\cite{Voutsinas:2019eyu}, introduce stochastic variations that
can significantly affect cross sections, threshold scans, and precision
measurements. Beamstrahlung arises when electrons or positrons are accelerated
in the intense electromagnetic field of the opposing bunch, leading to the
emission of energetic photons and a broadening of the luminosity
spectrum~\cite{Chen:1988zz,Schulte:1999tx}. Even in the absence of
beamstrahlung, the finite energy spread of the accelerator contributes
additional smearing, which must be carefully included in simulations. Accurate
modelling of these effects is essential for reliable predictions of both
inclusive rates and differential distributions, and for the interpretation of
precision measurements at future
machines~\cite{Schulte:1999tx,FCC:2018evy,Frixione:2022ofv}.

The current generation of lepton collider simulations typically models beam
dynamics using parametrized luminosity spectra that incorporate both
beamstrahlung and the intrinsic energy spread. Tools such as
Circe~\cite{Ohl:1996fi} and its successors provide realistic,
process-independent spectra for lepton collider experiments, allowing Monte
Carlo event generators to sample the effective collision energy on an
event-by-event basis. These parametrizations capture the essential features of
beamstrahlung and energy smearing for most phenomenological studies, and have
been widely validated against detailed beam-beam simulations. However, future
colliders will pose more stringent requirements: achieving sub-per-mille
precision in cross sections and differential distributions will demand more
accurate, flexible, and fully differential luminosity models. In particular,
generators will need to consistently account for correlations between
beamstrahlung photons and initial state radiation (ISR), incorporate
machine-specific non-Gaussian tails in the energy spectrum, and interface
seamlessly with multi-particle final states and higher-order matrix elements.
The development of such next-generation beam-dynamics simulations will be
crucial to fully exploit the precision potential of future lepton colliders.


\section{Electroweak Physics}


Electroweak (EW)
corrections~\cite{Schonherr:2017qcj,Denner:2019vbn,Frederix:2018nkq,Banerjee:2020rww}
are becoming an essential component of precision event simulation, with their
phenomenological impact strongly dependent on the collider environment. At
lepton colliders, ISR~\cite{Frixione:2022ofv} from the incoming beams plays a
particularly prominent role. The emission of multiple soft and collinear
photons leads to large logarithmic corrections of the form $\alpha^n
    \log^{2n}(s/m_e^2)$, which must be resummed to all orders to prevent them from
spoiling the accuracy of perturbative
expansions~\cite{Yennie:1961ad,Gribov:1972ri,Kuraev:1985hb,Jadach:1993yv}.
These effects significantly distort effective centre-of-mass energies on an
event-by-event basis, modifying both total rates and differential
distributions, and are especially critical in threshold scans and precision
measurements at the Z pole~\cite{Jadach:1993yv,Ward:1998ht,FCC:2018evy}, where
the steep energy dependence of the cross section amplifies the impact of even
small shifts in the effective collision energy.

The theoretical description of ISR at lepton colliders has reached a high level
of sophistication. Structure-function approaches~\cite{Skrzypek:1990qs,Cacciari:1992pz} treat
the incoming lepton as a source of photon radiation whose effect is encoded in
a universal radiator function, analogous in spirit to a parton distribution
function for the electron, and provide per-mille accuracy for inclusive
observables when carried to sufficient order. Amplitude-level exponentiation
techniques, developed within the Yennie--Frautschi--Suura (YFS)
framework~\cite{Yennie:1961ad,Jadach:1988gb}, offer a complementary and in some
respects more systematic approach: soft photon contributions are exponentiated
to all orders in $\alpha$, and hard photon corrections are added as finite
residuals order by order. Both approaches have been implemented in dedicated
tools, BHWIDE~\cite{Jadach:1995nk}, YFSWW~\cite{Jadach:2001mp},
KK2f~\cite{Jadach:1999vf}, which have served as the primary theoretical input
for the LEP precision programme and whose accuracy is well established for the
processes they were designed to describe. However, their integration into
modern, general-purpose event generators, particularly in combination with
higher-order matrix elements, multi-particle final states, and parton showers,
remains mostly incomplete with the exception of
\Sherpa~\cite{Price:2025fiu,Krauss:2022ajk} where the YFS resummation has been
mostly automated and matched to (N)NLO. Ensuring a consistent treatment of ISR
together with final-state radiation, interference effects, and realistic event
generation across a wide range of processes is therefore a key requirement for
future lepton collider studies, and one that the existing dedicated tools,
however precise, were not designed to fulfil on their own.

In recent years, there has also been renewed interest in formulating
initial-state radiation in terms of electron parton distribution functions
(PDFs)~\cite{Frixione:2019lga,Bertone:2019hks,Frixione:2021zdp,Frixione:2022ofv,Bertone:2022ktl,Stahlhofen:2025hqd,Schnubel:2025ejl,Garosi:2023bvq},
providing a framework that more closely parallels the treatment of QCD
radiation at hadron colliders. In this approach, the collinear logarithms
associated with photon emission are absorbed into scale-dependent electron and
photon densities, which obey DGLAP-type evolution equations including QED
effects. This formulation allows for a systematic inclusion of higher-order
logarithmic corrections and facilitates the consistent combination of ISR with
other perturbative ingredients, such as fixed-order calculations and parton
showers. Beyond pure QED evolution, modern developments have extended this
picture to include the full electroweak sector, leading to the notion of lepton
PDFs that incorporate photon, and in some formulations even weak boson,
components. While still less mature than their hadronic counterparts, these
approaches offer a promising route towards a unified treatment of initial-state
radiation across lepton and hadron collider environments. In particular, they
provide a natural language for describing photon-initiated processes and for
interfacing ISR with QCD evolution in mixed-coupling calculations, thereby
addressing some of the conceptual limitations of traditional structure-function
methods. The photon structure of the electron, relevant for resolved photon
processes at CLIC and to a lesser extent at ILC, introduces an additional layer
of complexity. At high energies, a significant fraction of the luminosity
involves interactions where one or both of the incoming leptons has radiated a
quasi-real photon that itself acts as a source of hadronic structure,
contributing processes of the type $\gamma\gamma \to \text{hadrons}$, $\gamma p
    \to X$ (with the proton replaced by the hadronic content of the photon), and
related
channels~\cite{Gluck:1991jc,Schuler:1995fk,Frixione:1993yw,Bertone:2017bme,Manohar:2016nzj,Manohar:2017eqh,Gehrmann:2025wbc,Abramowicz:1991yb}.
Further work is required to quantify their numerical impact in realistic
collider environments and to achieve a consistent integration into
general-purpose event generators. In particular, an important open task is the
combination of this approach with modern parton-shower frameworks, enabling a
fully differential description of multi-photon phase space, although initial
steps in this direction have already been
taken~\cite{CarloniCalame:2001ny,Belloni:2026vnv,Flower:2026byh,Skands:2020lkd}.

Beyond ISR, final-state radiation (FSR) from charged decay products introduces
its own set of complications. At the Z pole, QED corrections to hadronic and
leptonic Z decays are well understood at leading order and are implemented in
all generators, but the higher-order FSR contributions relevant for $e^+e^-$
precision, including $\mathcal{O}(\alpha^2)$
corrections~\cite{Bardin:1989tq,Montagna:1998kp,Bardin:1999yd,Arbuzov:2005ma,Baur:1997wa,Chen:2022dow,Freitas:2014hra}
to leptonic widths and the interference between radiation off different
final-state legs, are not uniformly available. In processes with hadronic final
states, the interplay between QED and QCD radiation from the same coloured
particle requires a coherent treatment that avoids double-counting between the
photon shower and the gluon shower while correctly reproducing the collinear
limits of both. This is a conceptually non-trivial problem: QED and QCD
radiation share the same infrared structure in the collinear limit, and a naive
superposition of independent showers misses interference terms that are
subleading in $\alpha_s$ or $\alpha$ individually but can be numerically
significant when both couplings are of comparable effective size, as they are
in several kinematic regions relevant for lepton colliders.

At higher orders, the mixed QCD--EW
corrections~\cite{Czarnecki:1996ei,Armadillo:2024nwk,Harlander:1998cmq,Dittmaier:2014qza,
Avdeev:1994db,Chen:2020xot,Bonciani:2020tvf,Cieri:2020ikq,Dittmaier:2020vra,Buccioni:2022kgy,
Bonciani:2021zzf,Delto:2019ewv,Bonciani:2019nuy,Buccioni:2020cfi,Behring:2020cqi,
Dittmaier:2024row,deFlorian:2018wcj,Armadillo:2022bgm,Heller:2020owb,Buonocore:2021rxx},
contributions of order $\alpha_s \alpha$ to processes involving both strong and
electroweak interactions, represent the frontier of fixed-order calculations
relevant for future lepton colliders. For processes where both QCD and EW
corrections contribute at the few-percent level, their product generates mixed
terms that are parametrically of order $\alpha_s \alpha \sim 10^{-4}$ but whose
numerical coefficient can be enhanced by large logarithms or by the proximity
to thresholds. For the $\alpha_s$ extraction from the hadronic Z width or from
event shapes, these mixed corrections enter the perturbative prediction at a
level that is not negligible relative to the precision goals, and their
systematic incorporation into a generator framework, rather than as a post-hoc
correction factor applied to an otherwise pure QCD or pure EW prediction, is an
open problem. The technical difficulty is substantial: mixed corrections
require two-loop calculations with both QCD and EW propagators, combined with a
consistent subtraction scheme that handles the overlapping infrared
singularities of the two interactions simultaneously.

An equally important challenge is the computation of pure two-loop electroweak
corrections, which become mandatory once experimental uncertainties reach the
sub-permille level anticipated at future lepton colliders. Such corrections are
enhanced by Sudakov logarithms of the form $\alpha^2 \log^n(s/M_W^2)$ at high
energies and can amount to several percent for multi-TeV processes, exceeding
the projected experimental precision if left unaccounted
for~\cite{Pozzorini:2001rs,Kuhn:2001hz,Denner:2003wi,
Denner:2006jr,Feucht:2004rp,Awramik:2003rn,
Awramik:2003ee,Awramik:2006uz,Dubovyk:2018rlg,Dubovyk:2019szj,Jantzen:2005az,Armadillo:2025mfx,Freitas:2025vax}. 
Even at the electroweak scale, two-loop
contributions enter precision observables such as the effective weak mixing
angle, the $W$-boson mass, and Z-pole pseudo-observables, where complete or
nearly complete calculations have been required to match the precision of LEP
and are indispensable for future facilities such as the ILC, CLIC, FCC-ee, and
CEPC. The calculation of generic
two-loop electroweak amplitudes remains technically demanding owing to the
presence of multiple mass scales, unstable particles, and overlapping infrared
and threshold singularities. Their automated incorporation into Monte Carlo
event generators, together with matching to higher-order QED radiation and
resummation frameworks, therefore represents one of the major theoretical
frontiers for precision phenomenology at future colliders.

Taken together, these considerations paint a picture of electroweak corrections
as a multi-layered problem in which no single ingredient dominates. The
hierarchy of effects, ISR resummation, FSR, ISR--FSR interference, resonance
treatment, non-factorisable corrections, mixed QCD--EW, EW Sudakov logarithms,
must all be addressed at a level commensurate with the precision goals of the
machine in question, and they must be addressed simultaneously in a coherent
generator framework rather than sequentially in isolated approximations. The
existing toolkit, while impressive, was built largely to serve the LEP and LHC
programmes, and its extension to meet the demands of FCC-ee, ILC, and CLIC will
require both new theoretical developments and substantial engineering effort.

\section{Perturbative QCD}
Fixed-order QCD corrections will play a crucial role at lepton colliders as
many of the key observables at lepton machines, ranging from hadronic event
shapes~\cite{Farhi:1977sg,Basham:1978bw,Dasgupta:2003iq,Larkoski:2014wba,Dasgupta:2013ihk,Banfi:2014sua,Georgi:1977sf,Gehrmann-DeRidder:2014hxk,Weinzierl:2009ms,Weinzierl:2009yz,Gehrmann-DeRidder:2009fgd,Catani:1992jc,Catani:1991kz,Gehrmann-DeRidder:2007vsv,Ellis:2010rwa,Salam:2010nqg,DelDuca:2016csb,Tulipant:2017ybb,Cesarotti:2020hwb}
and jet
rates~\cite{Ellis:1980wv,Gehrmann-DeRidder:2008qsl,Banfi:2016zlc,Baberuxki:2019ifp}
to heavy-flavour
production~\cite{Ball:2001pq,Dokshitzer:1995ev,Norrbin:2000uu,Krauss:2017wmx,Mele:1990yq}
and Higgs
decays~\cite{Wang:2023azz,Knobbe:2023njd,Perez:2015lra,Herzog:2017dtz,DelDuca:2015zqa,Mondini:2019gid,Fox:2025cuz,Coloretti:2022jcl,Gehrmann-DeRidder:2023uld,CampilloAveleira:2024fll},
are directly sensitive to QCD dynamics in the final state. As a result,
achieving the precision goals of future facilities requires a level of
theoretical control over QCD effects that is at least comparable to, and in
some cases exceeds, that reached at hadron colliders.

Measurements of Standard Model parameters, such as the strong coupling
constant, heavy-quark masses, and electroweak couplings, are expected to reach
unprecedented accuracy. In this context, fixed-order QCD calculations play a
central role in reducing theoretical uncertainties and providing systematically
improvable predictions. For many benchmark processes, NLO accuracy is no longer
sufficient, and NNLO corrections are required to match the projected
experimental precision.

Moreover, the clean experimental environment of lepton colliders enhances
sensitivity to subtle QCD effects that are often obscured at hadron machines.
Observables such as event shapes, jet substructure variables, and exclusive jet
cross sections probe multiple scales and are sensitive to both hard emissions
and the resummation of logarithmically enhanced contributions. Fixed-order
calculations provide the backbone for these predictions, defining the
normalisation and controlling the behaviour in regions dominated by hard,
well-separated partons. Their interplay with resummation and parton showers is
therefore essential for a reliable description across phase space

The past decades have seen remarkable progress in fixed-order QCD calculations,
with NNLO becoming the standard for many key
processes~\cite{Chawdhry:2019bji,Hartanto:2022qhh,Czakon:2021mjy,Alvarez:2023fhi,Chawdhry:2021hkp,Badger:2023mgf,Garbarino:2025bfg,Devoto:2024nhl,Buonocore:2023ljm,Buonocore:2022pqq,Catani:2022mfv,Biello:2024pgo,Kallweit:2020gcp,Chen:2022gpk,Chen:2019zmr,Gehrmann-DeRidder:2017mvr,Currie:2017eqf,Currie:2016bfm,Gehrmann-DeRidder:2016cdi,Gehrmann-DeRidder:2015wbt,Czakon:2012pz,Czakon:2012zr,Barnreuther:2012wtj}
and \NNNLO available for some
processes~\cite{Chen:2022xnd,Duhr:2020sdp,Chen:2022lwc,Cieri:2018oms,Billis:2019vxg,Chen:2021vtu,Baglio:2022wzu,Chen:2022vzo,He:2025hin,Duhr:2020seh,Duhr:2019kwi,Dreyer:2018qbw,Mondini:2019gid,Currie:2018fgr,Campbell:2023lcy,Chen:2025kez,Chen:2022cgv,Chen:2021isd}.
For lepton colliders we have a advantage that, generally speaking, most state
of the art advancements within perturbative QCD are first applied to the
simpler leptonic initial states, before being applied to processes with initial
state
hadrons.~\cite{Gehrmann-DeRidder:2004ttg,Chen:2025kez,Gehrmann-DeRidder:2007foh,Anastasiou:2004qd,Weinzierl:2009nz,Jakubcik:2022zdi,Garland:2001tf,Garland:2002ak}.

The challenge for MC generators is not to develop the fixed-order calculations
themselves, those are typically provided by dedicated codes, but their
consistent combination with the all-order QCD predictions of the corresponding
parton shower. Beyond fixed order, resummation of large logarithms plays an
important role at lepton colliders (where thrust, C-parameter and other event
shapes are precision observables) and at hadron colliders at small and large
transverse momenta. The interface between analytic resummation and numerical
parton showers, which both resum leading logarithms but in different schemes,
requires careful matching to avoid double-counting or the introduction of power
corrections not present in the analytic prediction. For FCC-ee observables
sensitive to leading-power QCD effects, achieving consistent NNLL+NNLO accuracy
in an MC framework is a concrete and pressing goal.

\NLOPS matching schemes such as \MCatNLO, \POWHEG, and their generalisations, are well-established, 
but \NNLOPS matching remains in active development. The complication is conceptual as much as technical: 
at NNLO, one encounters double-unresolved real emissions and two-loop virtual corrections whose interplay 
with the shower requires a coherent subtraction scheme that avoids both double-counting and gaps in phase-space coverage. 
Several approaches have been development such as \MINNLOPS~\cite{Monni:2019whf}, \UNNLOPS~\cite{Hoche:2014uhw}, \Geneva~\cite{Alioli:2015toa} and their systematic generalisation to more complex processes is ongoing work.

\section{Hadronization and Soft Physics}
Hadronization, which is the non-perturbative transition from coloured partons
to observable hadrons, is modelled phenomenologically in all generators, either
via string fragmentation~\cite{Artru:1974hr,Andersson:1983ia} or via cluster
hadronization~\cite{Gottschalk:1983fm,Chahal:2022rid,Bellm:2015jjp}. Both
approaches have been
tuned~\cite{Buckley:2009bj,Krishnamoorthy:2021nwv,LaCagnina:2023yvi,Bellm:2019owc,Ilten:2016csi,Lazzarin:2020uvv,Skands:2014pea,Iskauskas:2026rxi}
extensively to LEP and lower-energy $e^+e^-$ data, and they reproduce a wide
range of inclusive and semi-inclusive observables at those energies remarkably
well. However, the precision demands of FCC-ee, ILC, and CLIC expose
limitations that have been somewhat tolerable until now but will not remain so.

The most immediate concern is the role of hadronization corrections in
electroweak precision observables at future lepton colliders. Event shape
variables, such thrust, C-parameter, heavy jet mass, and others, are among the
most powerful probes of $\alpha_s$ at the Z pole~\cite{dEnterria:2022hzv}, and
their distributions receive non-perturbative corrections of order
$\Lambda_\text{QCD}/Q$ that are comparable in size to the higher-order
perturbative terms one is trying to extract. At LEP, these power
corrections~\cite{Dokshitzer:1997ew} were estimated using analytic dispersive
approaches and by comparing generator tunes, and they contributed a systematic
uncertainty to $\alpha_s$ determinations that was already competitive with the
perturbative uncertainty. For future lepton collider experiments, with
statistical uncertainties reduced some orders of magnitude, this situation
becomes acute: hadronization corrections are no longer a subdominant systematic
but potentially the dominant limitation on the extraction of $\alpha_s$ and
other precision quantities from event shapes. A systematic programme combining
analytic power correction calculations with constrained generator models,
validated across a wide range of observables simultaneously, is needed before
data arrives.

The top-quark mass measurement at the $t\bar{t}$ threshold brings hadronization
into contact with one of the most conceptually subtle problems in
QCD~\cite{Butenschoen:2016lpz}. The relationship between the MC mass parameter
used in generators and a well-defined short-distance mass schemes such as the
$\overline{\text{MS}}$ mass or the potential-subtracted mass, involves
non-perturbative contributions that are parametrically of order
$\Lambda_\text{QCD}$ and cannot be removed by perturbative matching alone. This
so-called renormalon~\cite{Beneke:1998ui} ambiguity means that extracting a
theoretically clean top mass from a threshold scan requires not only a precise
generator description of the lineshape, but a quantitative understanding of the
hadronization corrections to the threshold cross section and their scheme
dependence. The size of these corrections is large compared to the 10–20 MeV
experimental precision target, and reducing their uncertainty to an acceptable
level is an open theoretical and modelling problem.

Heavy-flavour production and
fragmentation~\cite{Mele:1990cw,Cacciari:1998it,Kretzer:2000yf,Frixione:1997ma,ALEPH:2001pfo}
represent another area where lepton collider precision exposes the limits of
current models. The fragmentation of b and c quarks into hadrons is described
in generators through a combination of perturbative evolution and
non-perturbative fragmentation functions, tuned primarily to LEP measurements
of B-meson spectra and charm production rates. Enormous samples of B-mesons,
D-mesons, baryons, and their excited states will allow measurements of
fragmentation functions~\cite{Collins:1992kk} with a statistical precision that
far exceeds the accuracy of current non-perturbative models. The modelling of
the b-quark fragmentation function near the kinematic endpoint, the production
rates of orbitally and radially excited heavy-flavour states, and the relative
contributions of different B-hadron species are all areas where present
generator descriptions are known to be approximate. Improving them requires
both better theoretical input from lattice QCD, SCET-based fragmentation
function calculations, and analytic approaches to endpoint resummation and a
more flexible parametrization within the generator framework that can be
constrained by data.

Colour
reconnection~\cite{Sjostrand:1993hi,Christiansen:2015yca,Gieseke:2012ft,Bellm:2019wrh}
and related collective effects are increasingly recognised as important sources
of systematic uncertainty in precision measurements at lepton colliders. In
$e^+e^- \to W^+W^-$ events, the two W bosons decay hadronically with a spatial
overlap that, at the parton shower level, allows gluons from one W decay to be
colour-connected to quarks from the other. This colour reconnection between the
two independently evolving colour singlets shifts the reconstructed W invariant
mass and introduces a systematic uncertainty in \MW measurements that is not
present in perturbative theory and must be modelled and estimated at the
generator level. Current models of colour reconnection differ in their
predictions for this effect by amounts that are significant relative to the
precision targets of future colliders, and a principled theoretical framework,
rather than a collection of phenomenological models with unconstrained
parameters, is needed to make progress. Related effects, including
string-string interactions~\cite{Bierlich:2014xba,Bierlich:2020naj} (string
shoving) in dense partonic environments and the Bose-Einstein
correlations~\cite{Lonnblad:1997kk,Lonnblad:1995mr} between identical pions
from different W decays, are similarly unresolved and similarly important for
the \MW measurement programme.

The modelling of the underlying hadronic
activity~\cite{Sjostrand:1987su,Sjostrand:2004pf} in $e^+e^-$ events, including
secondary interactions, photon-induced
processes~\cite{Walsh:1973mz,Klasen:2002xb}, and the transition region between
perturbative and non-perturbative dynamics, requires careful attention at CLIC
energies where the two-photon background from beamstrahlung becomes a
significant overlay. At 3 TeV, the hadronic activity generated by $\gamma\gamma
    \to \text{hadrons}$ events piling up on hard interactions resembles, in some
respects, the underlying event at a hadron collider, but with a different
colour structure and a softer transverse momentum distribution. The soft
physics models used to describe this background, themselves derived from fits
to low-energy photon-photon data, are being extrapolated far beyond their
validation range, and the uncertainties in this extrapolation will affect jet
energy measurements, missing energy estimates, and the modelling of forward
activity in essentially every physics analysis.

More broadly, the entire hadronization model at lepton colliders rests on a
foundation of LEP data collected some thirty years ago. Tunes performed to
those data describe a specific kinematic regime and a specific mix of quark
flavours set by the Z decay fractions; they are not guaranteed to extrapolate
correctly to the very different conditions of FCC-ee's WW and ZH stages, where
the colour topology and the typical parton energies differ substantially from
the Z pole. A systematic study of the sensitivity of precision observables at
each FCC-ee energy stage to hadronization model parameters, analogous to the
underlying event and fragmentation sensitivity studies performed for LHC
analyses, is needed to identify where the extrapolation is safe and where
dedicated re-tuning or theoretical improvement is required.

\section{Conclusion}

Monte Carlo event generators have become an indispensable component of modern
collider physics, providing the only practical bridge between first-principles
theory and realistic experimental measurements. Their success at the LHC
reflects decades of sustained theoretical and technical development, supported
by continuous validation against data. However, the physics goals of future
collider programmes, in particular high-precision lepton colliders place
demands on generators that go well beyond this established baseline.

Across all areas discussed in this work, electroweak corrections, initial-state
radiation, beam dynamics, perturbative QCD, and non-perturbative modelling, a
common picture emerges: no single limitation dominates, but rather a network of
interdependent effects must be controlled simultaneously at an unprecedented
level of precision. In many cases, the required ingredients already exist in
isolation, for example in dedicated high-precision tools or fixed-order
calculations, but their consistent combination within general-purpose event
generators remains incomplete. Achieving this synthesis, while maintaining
flexibility, efficiency, and full event exclusivity, represents one of the
central challenges for the field.

At the same time, the scale of future simulation campaigns will necessitate
substantial advances in computational performance and workflow design. The
integration of higher-order calculations, resummation, and complex beam and
detector effects must be achieved without prohibitive increases in computing
cost, motivating the exploration of new algorithmic strategies, parallel
architectures, and machine learning techniques.

Ultimately, the development of next-generation MC generators is not a purely
technical task, but a core component of the physics programme itself. The
precision targets of future colliders cannot be realised without corresponding
advances in theoretical predictions at the fully differential level. This
requires a coordinated effort across theory, phenomenology, and experiment,
building on the lessons of the LHC while addressing the qualitatively new
challenges posed by future machines. The work outlined here represents only a
snapshot of an active and rapidly evolving field, but it highlights clearly
that significant progress, both conceptual and practical, will be required in
the coming years to fully exploit the potential of the next generation of
collider experiments.

\section{Acknowledgements}
The author would like to thank the editors for their invitation to submit this
contribution. The work of A.P. is supported by grant No. 2023/50/A/ST2/00224 of
the National Science Centre (NCN), Poland. 

\bibliography{sn-article}
\end{document}